# A simple numerical scheme for the 2D shallow-water system

Jie Hu, Graduate Research Student, *The National Hydraulics and Environmental Laboratory* ***LNHE*** *(**L**aboratoire **N**ational d'**H**ydraulique et **E**nvironnement), R&D, EDF, 6 Quai Watier, 78400 Chatou, France*

*E-mail address*: e0013640@u.nus.edu (author for correspondence)

*Presently at: 3 Engineering Drive 2, NUS, 117578, Singapore*

## Abstract

This paper presents a simple numerical scheme for the two dimensional Shallow-Water Equations (SWEs). Inspired by the study of numerical approximation of the one dimensional SWEs Audusse et al. (2015)**,** this paper extends the problem from 1D to 2D with the simplicity of application preserves. The new scheme is implemented into the code TELEMAC-2D [tel2d, 2014] and several tests are made to verify the scheme ability under an equilibrium state at rest and different types of flow regime (i.e., fluvial regime, transcritical flow from fluvial to torrential regime, transcritical flow with a hydraulic jump). The sensitivity analysis is conducted to exam the scheme convergence.

## 1. Introduction

The SWEs have been proposed by Saint-Venant (1871) to model flows in a channel. Nowadays, they are used to model flows in a wide variety of physical phenomena, such as: overland flow, flooding, dam breaks, tsunami (e.g. Esteves et al. 2000, Caleffi et al. 2003, Valiani et al. 2002, and Kim et al. 2007). These equations are a time-dependent two-dimensional system of non-linear partial differential equations of hyperbolic type.

In real situations (realistic geometry, sharp spatial), there is little hope to solve explicitly the SWEs, i.e. to produce analytic formula for the solutions. Therefore, it is necessary to develop specific numerical methods to compute approximate solutions of SWEs (e.g. Toro 1999,

Bouchut, 2004, and LeVeque, R.J., 2002). Implementation of any of such methods raises the question of the validation of the code.

Validation is a necessary step to check if a model (the numerical methods) suitably describes the considered phenomena. There exists at least three complementary types of numerical tests to ensure a numerical code is relevant for the considered systems of equations. First, one can produce convergence or stability results (e.g. by refining the mesh). This validates only the numerical method and its implementation. Second, approximate solutions can be matched to analytic solutions available for some simplified or specific cases. Finally, numerical results can be compared with experimental data, produced indoor or outdoor. This step should be done after the previous two; it is the most difficult one and must be validated by a specialist of the domain. This paper focuses on the first two steps.

A simply implementary scheme for 1D SWEs is provided in Audusse et al. (2015), and this scheme is proved to be accurate and robust on several typical test cases. Enlightened by Audusse et al. (2015)'s work, the present paper describes a numerical scheme for the 2D SWEs to study the free surface shallow flows.

The paper is organized as follows: the general mathematical model is described in section 2. In section 3, the property of rotational invariance of SWEs is applied to split the governing equations into $x$ direction, thus simplifying the problem. The classic Riemann solver is reviewed in section 4 and a Godunov-type finite volume scheme is derived for the augmented 1D SWEs. A simple treatment of the source term is adopted from Audusse et al. (2015), in section 5. Several test cases are conducted to exam the ability of the new scheme under the equilibrium state and several types of flow regime, in section 6.

## 2. Mathematical model

The SWEs can be deduced from Navier-Stokes equations for an incompressible fluid by making the hypothesis of hydrostatic pressure, uniform velocities along the vertical direction. For inviscid flow, the model can be written in its conservative form as follows

$$\frac{\partial \boldsymbol{U}}{\partial t}+\frac{\partial}{\partial x}F(\boldsymbol{U})+\frac{\partial}{\partial y}G(\boldsymbol{U})=S_s(\boldsymbol{U})+S_f(\boldsymbol{U});\ (x,y)\in\Omega, t\geq 0 \qquad \text{Eq. 1}$$

$$\boldsymbol{U}=\begin{pmatrix} h \\ hu \\ hv \end{pmatrix},\qquad F(\boldsymbol{U})=\begin{pmatrix} hu \\ hu^2+\frac{1}{2}gh^2 \\ huv \end{pmatrix},\qquad G(\boldsymbol{U})=\begin{pmatrix} hv \\ hvu \\ hv^2+\frac{1}{2}gh^2 \end{pmatrix},$$

$$S_s(\boldsymbol{U})=\begin{pmatrix} 0 \\ -gh\frac{\partial b}{\partial x} \\ -gh\frac{\partial b}{\partial y} \end{pmatrix},\quad S_f(\boldsymbol{U})=\begin{pmatrix} 0 \\ -g\frac{\mathcal{N}^2\sqrt{u^2+v^2}}{h^{4/3}}hu \\ -g\frac{\mathcal{N}^2\sqrt{u^2+v^2}}{h^{4/3}}hv \end{pmatrix},\qquad \mathcal{N}=coef\ of\ Manning\ friction$$

where $b(x,y)$ is a smooth topography, $g$ refers to the gravitational acceleration, $h$ is the water height, $u$ and $v$ are horizontal and vertical velocity components. $h$, $u$ and $v$ are functions of time $t$ and space $x, y$.

## 3. Rotational invariance

We first consider the homogeneous (no source terms) time-dependent two-dimensional SWEs.

$$\boldsymbol{U}_t+F(\boldsymbol{U})_x+G(\boldsymbol{U})_y=\boldsymbol{0} \qquad \text{Eq. 2}$$

where $\boldsymbol{U}_t$ is $\frac{\partial \mathbf{U}}{\partial \mathrm{t}}$, $F(\boldsymbol{U})_x$ and $G(\boldsymbol{U})_y$ are $\frac{\partial}{\partial x}F(\boldsymbol{U})$ and $\frac{\partial}{\partial y}G(\boldsymbol{U})$ in Eq. 1. Following the rotational invariance property of the SWEs by Toro (2001), the two-dimensional problem can be reduced to augmented one-dimensional:

$$\boldsymbol{U}_t^*+F(\boldsymbol{U}^*)_x=\boldsymbol{0} \qquad \text{Eq. 3}$$

where $\boldsymbol{U}^*=[h,h\tilde{u},h\tilde{v}]^T$, $\tilde{u},\tilde{v}$ are the velocity components after the angular rotational by a rotation matrix $T=\begin{bmatrix}1 & 0 & 0\\ 0 & cos\theta & sin\theta \\ 0 & -sin\theta & cos\theta\end{bmatrix}$. The new velocities are related with the original ones by $\tilde{u}=ucos\theta+vsin\theta$, $\tilde{v}=-usin\theta+vcos\theta$. The tilde $\widetilde{(.)}$ and (*) is dropped later for brevity.

## 4. The Riemann Problem and the Godunov Flux

Our concern is about solving numerically the general initial-boundary value problem (IBVP) for the augmented one-dimension SWEs in section 3.

$$\boldsymbol{U}_t + F(\boldsymbol{U})_x = \boldsymbol{0}$$

$$\boldsymbol{U}(x,0) = \begin{cases} U_L, & x_l < x < 0 \\ U_R, & x_r > x > 0 \end{cases} \qquad \text{Eq. 4}$$

with given states $U_L = [h_L, h_L u_L, h_L v_L\ ]^T$, and $U_R = [h_R, h_R u_R, h_R v_R\ ]^T$, the subscript *L(R)* mean *Left(Right)* side, this forms a Riemann problem.

Utilising Godunov-type methods based on the explicit conservative formula

$$\boldsymbol{U}_i^{n+1} = \boldsymbol{U}_i^n - \frac{\Delta t}{\Delta x}\left[F_{i+\frac{1}{2}} - F_{i-\frac{1}{2}}\right],\ \forall\, i \in \mathbb{Z} \qquad \text{Eq. 5}$$

where $\boldsymbol{U}_i^n$ is the solution of the Eq. 4 in cell $C_i = \left[x_{i-\frac{1}{2}},\ x_{i+\frac{1}{2}}\right]$ centred at point $x_i$ at time $t^n$. *i* and *n* are space and time index. $x_{i-\frac{1}{2}}, x_{i+\frac{1}{2}}$ are cell interfaces, space step $\Delta x = x_{i+\frac{1}{2}} - x_{i-\frac{1}{2}}$ and time space $\Delta t = t^{n+1} - t^n$. $F_{i+\frac{1}{2}}$ is numerical flux at cell interface $x_{i+\frac{1}{2}}$. The vectors of conserved variable and fluxes are $\boldsymbol{U} = \begin{pmatrix} h \\ hu \\ hv \end{pmatrix}$, $F(\boldsymbol{U}) = \begin{pmatrix} hu \\ hu^2 + \frac{1}{2}gh^2 \\ huv \end{pmatrix}$.

The general structure of the approximate solution of the Riemann problem for the augmented one-dimensional is depicted in Fig 1. The value of the solution along $x/t = 0$ corresponds to the *t*-axis and is the value required for the computation of the Godunov flux.

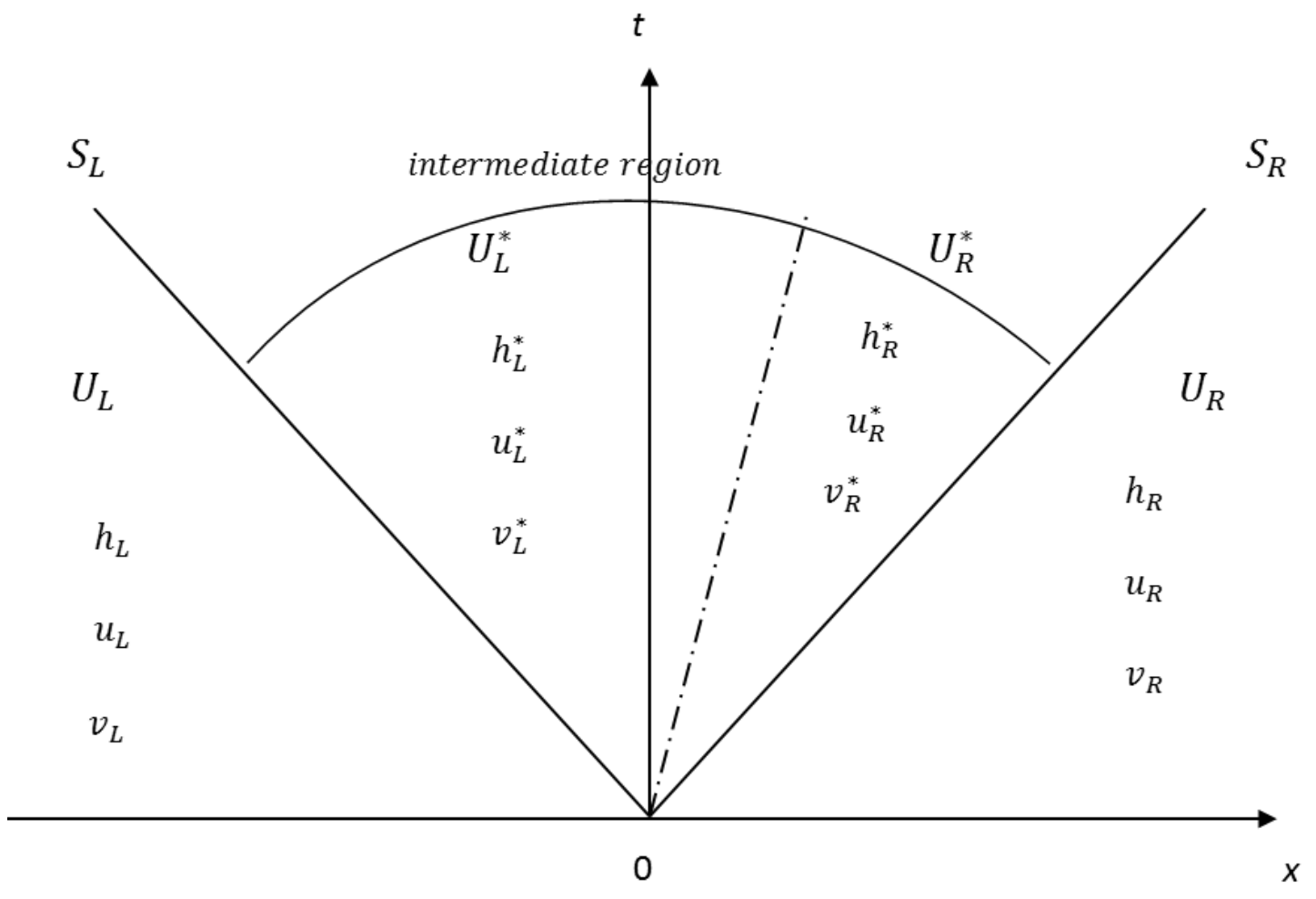

Fig 1 Structure of the approximate solution of the Riemann problem for the *x*-split homogenous two-dimensional SWEs. There are three approximate wave families $S_L$, $S_R$ and $S^*$ associated with the proper eigenvalues of Eq. 4.

### 4.1 The HLLC Riemann solver

Harten et al. (1983) suggested a way of solving the Riemann problem approximately by finding an approximation to the numeral flux $F_{i+\frac{1}{2}}$ (flux through the interface). The mathematical bases of the approach are given in Toro (1999). Assuming all wave speed estimates are available, the HLLC numerical flux is shown as follows

$$F_{i+\frac{1}{2}}^{hllc} = \begin{cases} F_L, & if \quad 0 \leq S_L, \\ F_L^*, & if \quad S_L \leq 0 \leq S^*, \\ F_R^*, & if \quad S^* \leq 0 \leq S_R, \\ F_R, & if \quad 0 \geq S_R \end{cases} \qquad \text{Eq. 6}$$

where

$$F_L^* = \quad F_L + S_L(U_L^* - U_L) \quad \& \quad F_R^* = \quad F_R + S_R(U_R^* - U_R) \qquad \text{Eq. 7}$$

The states $U_L^*$, $U_R^*$ are given by

$$U_K^* = \quad h_K \left(\frac{S_K - u_K}{S_K - S^*}\right) \begin{bmatrix} 1 \\ S^* \\ v_K \end{bmatrix}, \quad (K = L, R) \qquad \text{Eq. 8}$$

For determining the numerical flux, the three approximate wave speeds $S_L$, $S_R$ and $S^*$should *a priori* be known.

#### 4.1.1 Estimation of speeds $S_L$, $S_R$ and $S^*$

The determination of numerical flux in Eq. 6 requires the pre-known estimated wave speeds. Toro (2001) suggests the following choice of wave speeds that can lead to accurate and robust scheme:

$$S_{L(R)} = u_{L(R)} - a_{L(R)} q_{L(R)} \qquad \text{Eq. 9}$$

where $a_K = \sqrt{g h_K}$ and $q_K$ $(K = L, R)$ is given by

$$q_K = \begin{cases} \sqrt{\frac{1}{2}\left[\frac{(h^* + h_K) h^*}{h_K^2}\right]} & if \quad h^* > h_K \\ 1 & if \quad h^* \leq h_K \end{cases} \qquad \text{Eq. 10}$$

here $h^*$ is an estimate for the exact solution for $h$ in the intermediate region. Similar choices can also be found in Bello et al. (2007).

4.1.2 Estimation of water depth $h^*$ and particle velocity $u^*$ in the intermediate region

The primitive variables in intermediate region are *a prior* needed to determine wave speeds in sub-section 4.1.1. In Toro (2001), a new scheme solver is given for $h^*$and $u^*$.

$$\left.\begin{aligned} h^* &= \tfrac{1}{2}(h_L + h_R) - \tfrac{1}{4}(u_R - u_L)(h_L + h_R)/(a_L + a_R) \\ u^* &= \tfrac{1}{2}(u_L + u_R) - (h_R - h_L)(a_L + a_R)/(h_L + h_R) \end{aligned}\right\} \quad \text{Eq. 11}$$

This new solver has a simple form. It can deal very well with situations involving very shallow water and is found to be very robust in dealing with shock waves Toro (1995).

## 4.2 Including of the topography

The previous section (4.1) elaborates the HLLC scheme for the homogeneous SWEs. In this section, our scheme is extended to include the topography term. And Eq. 4 becomes

$$\boldsymbol{U}_t + F(\boldsymbol{U})_x = \begin{pmatrix} 0 \\ -gh\partial_x b \\ 0 \end{pmatrix} \quad \text{Eq. 12}$$

4.2.1 An uphill bottom $\partial_x b < 0$

If the topography is not flat and the left bottom height $b_1$ is less higher than the right bottom height $b_2$, which means the left water depth $h_L$ is greater than the right water depth $h_R$, if the free surface is at the same elevation. Then the first approximation of the height in intermediate region $h^*$ is smaller than $h_L$, but bigger than $h_R$ ($h_L > h^* > h_R$). We deduce that $S^*$is positive. Left intermediate region $U_L^*$ in Fig. 1 is divided into two zones denoted as $U_1$ and $U_2$ due to the influence of the topography. This is equivalent to introduce a velocity whose value is zero at *t*-axis, shown in Fig 2.

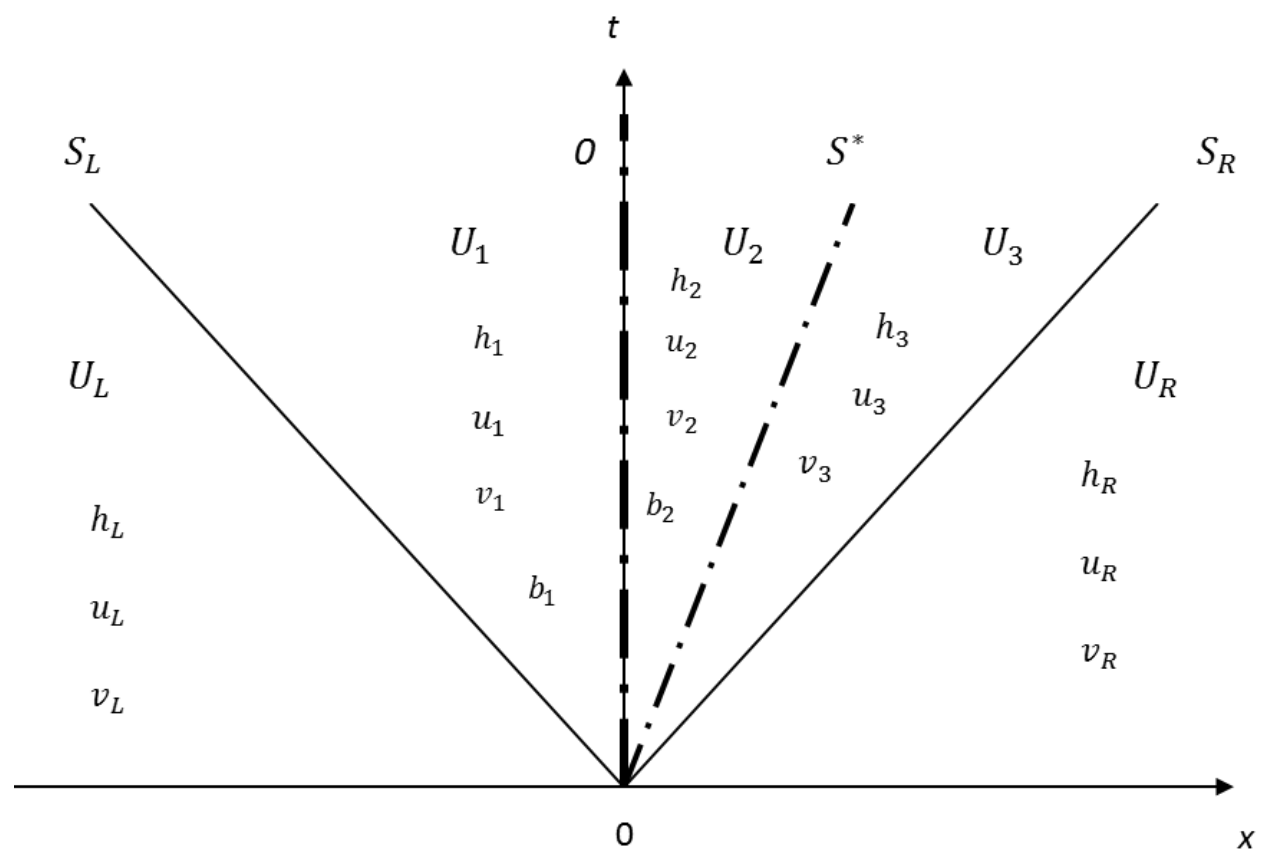

Fig 2. Structure of the approximate solution of the Riemann problem for the *x*-split two-dimensional SWEs with uphill topography. There are four approximate waves with speeds $S_L$, 0, $S^*$ and $S_R$, while 0 is induced by the effect of topography.

Integrating Eq. 12 on volume $[x_l, x_r]$x$[0, t]$ and deploy the first two components (as flux is along *x*-axis, no impact on third component), we can have:

$$h_R u_R - h_L u_L = S_L(h_1 - h_L) + S^*(h_3 - h_2) + S_R(h_R - h_3) \quad \text{Eq. 13}$$

$$\left(h_R u_R^2 + \frac{g h_R^2}{2}\right) - \left(h_L u_L^2 + \frac{g h_L^2}{2}\right) + g\Delta x\{h\partial_x b\} = S_L(h_1 u_1 - h_L u_L) + S^*(h_3 u_3 - h_2 u_2) + S_R(h_R u_R - h_3 u_3) \quad \text{Eq. 14}$$

where $\{h\partial_x b\}$ stands for a consistent approximation of the source term $h\partial_x b$ which will be precised later on. In order to close this system, two relations are missing and we suggest to impose:

- Hydraulic balance between states $U_1$ and $U_2$.

$$\begin{cases} h_1 u_1 = q_1 = h_1 u_1 = q_2 = q^* \\ h_1 + b_1 = h_2 + b_2 \end{cases} \quad \text{Eq. 15}$$

- The property of shear wave ($S^*$) from state $U_2$ to $U_3$.

$$\begin{cases} h_2 = h_3 \\ u_2 = u_3 \\ v_2 \neq v_3 \end{cases} \quad \text{Eq. 16}$$

By solving the Eq. 13, 15 and 16, we define the water heights at the left and the right side of *t*-axis as

$$h_1 = h_{HLLC} + \frac{S_R}{S_R - S_L}\Delta b \quad \text{Eq. 17}$$

$$h_2 = h_{HLLC} + \frac{S_L}{S_R - S_L}\Delta b \quad \text{Eq. 18}$$

where $\Delta b = b_2 - b_1$ and $h_{HLLC} = \frac{S_R h_R - S_L h_L - (h_R u_R - h_L u_L)}{S_R - S_L}$ is the intermediate water height associated to the HLLC solver (Chapter 10, Toro 1999).

By solving the Eq. 14, 15 and 16, we deduce the discharge at the left and the right side of *t*-axis as:

$$q^* = h_L^* u_L^* = h_R^* u_R^*,$$

$$q^* = q_{HLLC} - \frac{g}{S_R - S_L}\Delta x\, \{h\partial_x b\}, \qquad \text{Eq. 19}$$

with $q_{HLLC} = \frac{S_R h_R u_R - S_L h_L u_L - \left(h_R u_R^2 + \frac{g h_R^2}{2}\right) + \left(h_L u_L^2 + \frac{g h_L^2}{2}\right)}{S_R - S_L}$ is the intermediate discharge involved in the HLLC scheme (Chapter 10, Toro 1999).

From (17), (18) and (19), after some algebraic manipulation, we are able to define two numerical fluxes $F^L$ and $F^R$:

$$\begin{cases} F^L(\boldsymbol{U}_L, \boldsymbol{U}_R, b_1, b_2) = F_L^*(\boldsymbol{U}_L, \boldsymbol{U}_R) + \begin{pmatrix} \frac{S_L S_R}{S_R - S_L}\Delta b \\ -\frac{S_L g}{S_R - S_L}\Delta x\,\{h\partial_x b\} \end{pmatrix} \\ F^R(\boldsymbol{U}_L, U_R, b_1, b_2) = F_L^*(\boldsymbol{U}_L, \boldsymbol{U}_R) + \begin{pmatrix} \frac{S_L S_R}{S_R - S_L}\Delta b \\ -\frac{S_R g}{S_R - S_L}\Delta x\,\{h\partial_x b\} \end{pmatrix} \end{cases} \qquad \text{Eq. 20}$$

where $F_L^*$ is the HLLC flux in Eq. 6.

Hence, the numerical flux for SWEs with uphill bottom is:

$$F_{i+\frac{1}{2}}^{hllc}(\boldsymbol{U}_L, \boldsymbol{U}_R) = \begin{cases} F_i, & if \quad 0 \le S_L, \\ F_{i+\frac{1}{2}}^L = F_L^*(\boldsymbol{U}_i, \boldsymbol{U}_{i+1}) + \begin{pmatrix} \frac{S_L S_R (b_{i+1} - b_i)}{S_R - S_L} \\ -\frac{S_L g \Delta x\,\{h\partial_x b\}}{S_R - S_L} \end{pmatrix}_{i+1/2}, & if \quad S_L \le 0 \le S^*, \\ F_{i+\frac{1}{2}}^R = F_L^*(\boldsymbol{U}_i, \boldsymbol{U}_{i+1}) + \begin{pmatrix} \frac{S_L S_R (b_{i+1} - b_i)}{S_R - S_L} \\ -\frac{S_R g \Delta x\,\{h\partial_x b\}}{S_R - S_L} \end{pmatrix}_{i+1/2} & if \quad S_L \le 0 \le S^*, \\ F_{i+1}, & if \quad 0 \ge S_R \end{cases} \qquad \text{Eq.21}$$

It should note that similar expression can also be found in Audusse et al. (2015).

### 4.2.2 An downhill bottom $\partial_x b > 0$

Analogous with 4.2.1, for the topography with the left bottom height $b_1$ higher than the right bottom height $b_2$, it can be deduced that $S^*$ is negative, it has been found that the numerical flux for SWEs with downhill bottom has the same expressions as in Eq. 21 except the HLLC flux $F_L^*$ is replaced by $F_R^*$ in Eq. 6.

# 5 Treatment of the source term $\{h\partial_x b\}$

An adopted method is from Audusse et al. (2015) to discrete source term as follows

$$\{h\partial_x b\} = \begin{cases} \frac{h_L+h_R}{2\Delta x} min(h_L, \Delta b), & if \quad \Delta b \geq 0, \\ \frac{h_L+h_R}{2\Delta x} \max(-h_R, \Delta b), & if \quad \Delta b < 0, \end{cases} \qquad \text{Eq. 22}$$

This treatment can preserve the lake at rest in the case of a wet-dry transition or the case of a dry-wet transition.

# 6 Numerical results

The convergence of the new numerical scheme is first examined on two flow regimes (a fluvial flow and a transcritial flow). This validates the numerical method and its implementation. The behaviour of our new scheme is then examined on several test cases to check if the approximate solutions are accurate. These test cases include: an equilibrium state with a bump bottom, a steady flow over this bump in two regimes (fluvial and transcritical).

## 6.1 Mesh sensitivity study

To ensure the convergence of our new scheme, three different meshes are built for a rectangular topography. This topography is discretized by $N$ triangular elements, where $N$ are 968, 1925 and 3839, respectively (see Fig 3). The free software Bluekenue is used for creating these meshes. And more information of the rectangular topography is in section 6.2.

The relative error norms are defined as following:

$$||e_1(N)|| = \frac{\int_0^L |H_N(x) - H_{exact}(x)| dx}{L} \qquad \text{Eq. 23.1}$$

$$||e_2(N)|| = \frac{\sqrt{\int_0^L (H_N(x) - H_{exact}(x))^2 dx}}{L} \qquad \text{Eq. 23.2}$$

$$||e_1(N)|| = \max(H_N(x) - H_{exact}(x)) \qquad \text{Eq. 23.3}$$

with $H_N(x)$ the free surface (water depth $h$ plus bottom elevation $b$) profile from the new scheme with $N$ cell elements used and $H_{exact}(x)$ is the free surface profile from exact solution as the reference result with the same cell elements. $L$ is the total length of simulated domain and $dx$ is the space step. The simulations are performed with different cell element numbers (i.e., 968, 1925, 3839) to compute these error norms. The norms are identical for the discharge. The sensitivity study results for a fluvial flow and a transcritical flow and are depicted in Fig 4 and 5, respectively. The detailed conditions about these two flow regimes are elaborated in the next section 6.3 & 6.4.

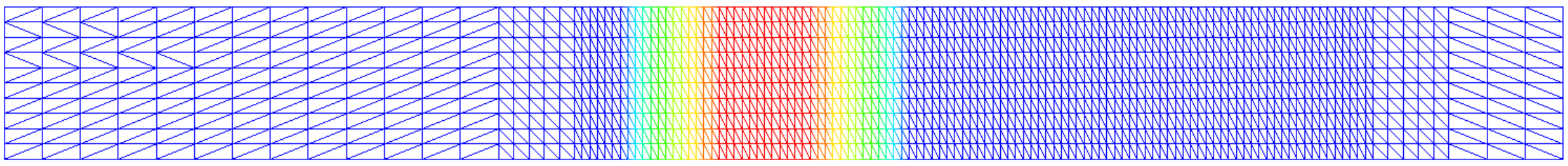

Fig 3 Top view of the topography discretized by 968 triangular elements, the refined area is with more intensive elements, different colours represent different elevations of the topography.

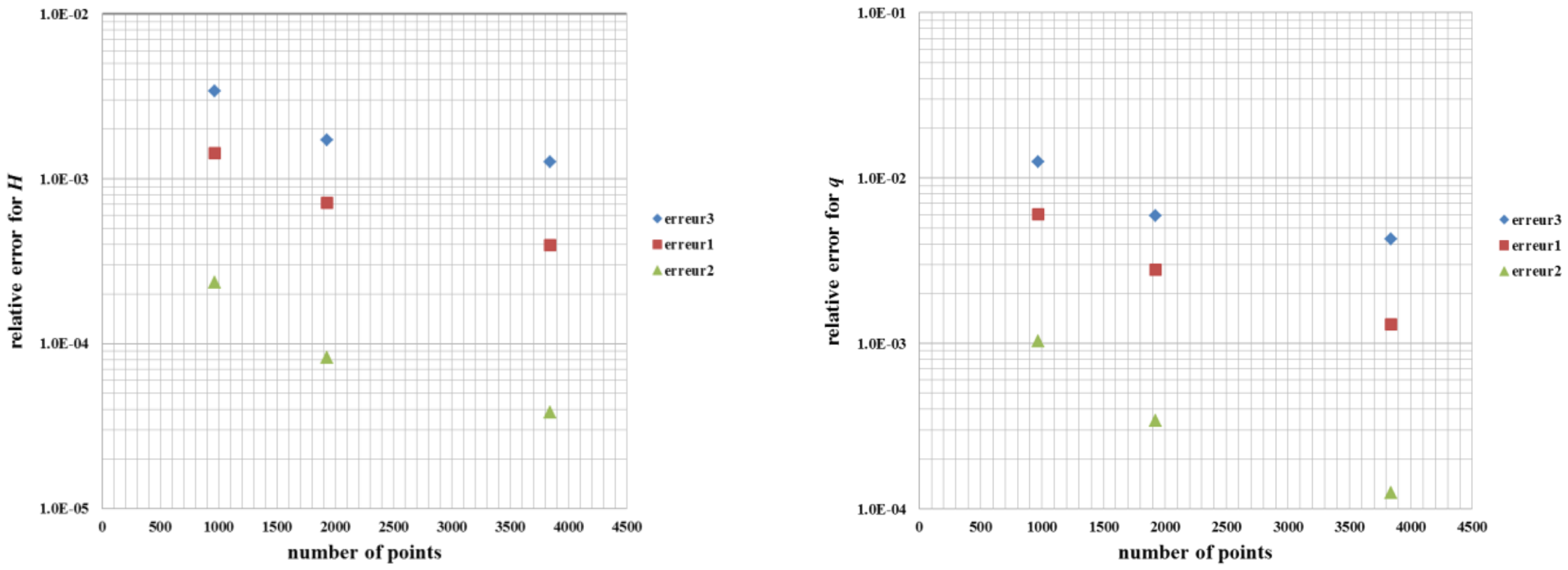

Fig 4 Fluvial flow: Comparison of orders of error for the free surface (left) and the discharge (right) of new scheme and exact solution

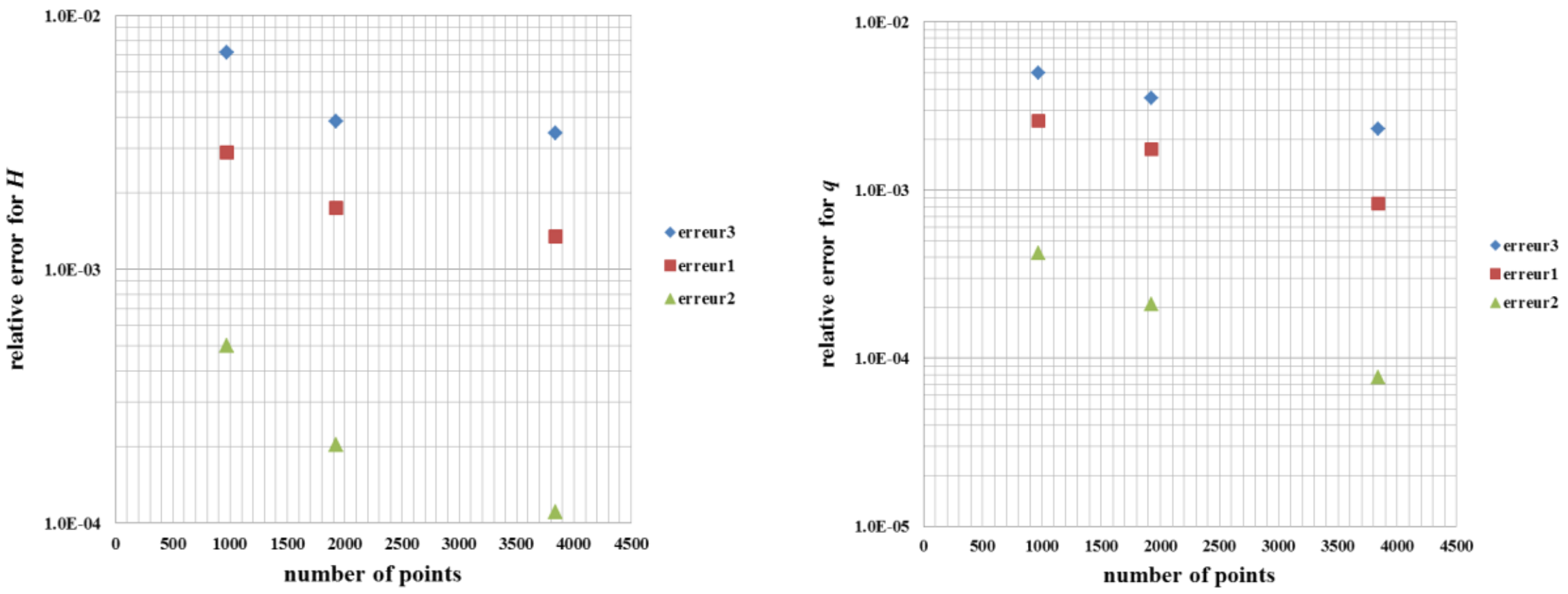

Fig 5 Transcritical flow: Comparison of orders of error for the free surface (left) and the discharge (right) of new scheme and exact solution

From Fig 4 and 5, it can be seen that the orders of error decrease monotonously with the increasing cell element numbers for different flow regimes. The convergence of our new scheme can thus be guaranteed.

### 6.2 Equilibrium steady state at rest

The aim of this test is to verify the scheme for equilibrium steady state. The topography is flat at both left and right sides except there is a bump in the middle. The length of the topography is 20.5 *m*. And the topography is discretized by 968 elements (Fig 3). This geometry is adopted for all the following tests. The free surface is initially constant with a height 1.8 *m* which submerges all the topography. At the right side, we impose a 1.8 *m* water depth which is equal to the initial condition, and at the left side, a zero discharge is imposed. In this way, we can ensure the equilibrium state. The time step is 0.01 *s*, and the total number of time steps is 10000 which is long enough to ensure the steady state. After the simulation, the result of free surface is shown in Fig 6.

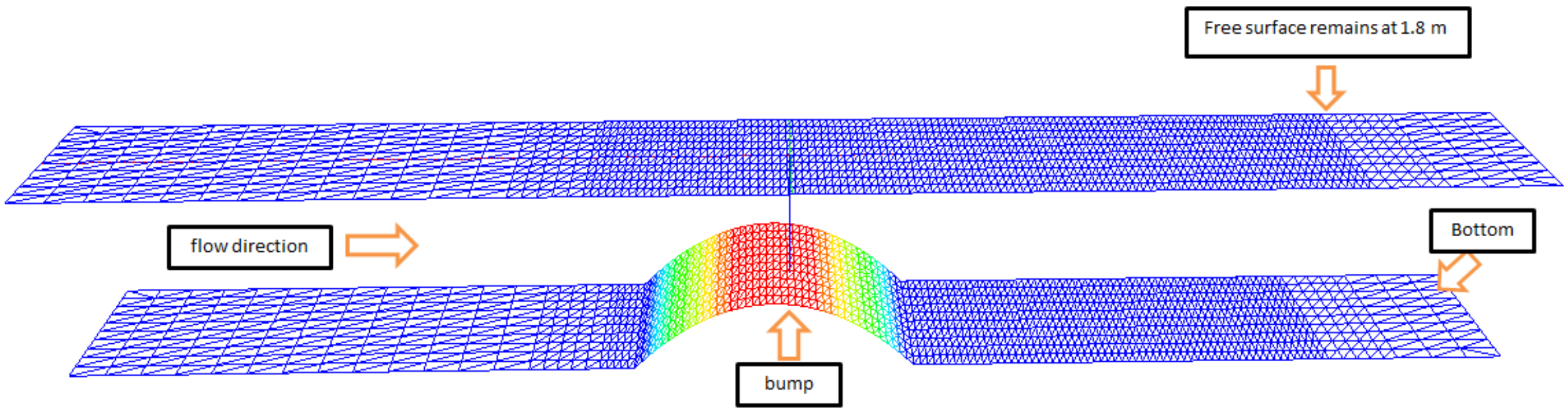

Fig 6 A front view of equilibrium steady state at rest over a bump

The Fig 6 is generated in TELEMAC-2D of the Telemac-Mascaret modelling system. It is noted that the free surface remains stable since the simulation starts. We can thus believe that our scheme can assure the equilibrium steady state at rest. Next, we exam our scheme in non-equilibrium steady states which consist of three types of situation: fluvial regime, transcritical flow from fluvial regime to torrential regime, transcritical flow with a hydraulic jump. The results calculated by our new scheme are compared quantitatively with exact solution.

### 6.3 Fluvial regime

The initial condition is the same as that in 6.2. A 1.8 *m* water depth is imposed at the right side and a discharge of 8.85889 $m^3/s$ is imposed at the left side. After the flow is at steady state, the free surface and discharge is depicted in Fig 6.

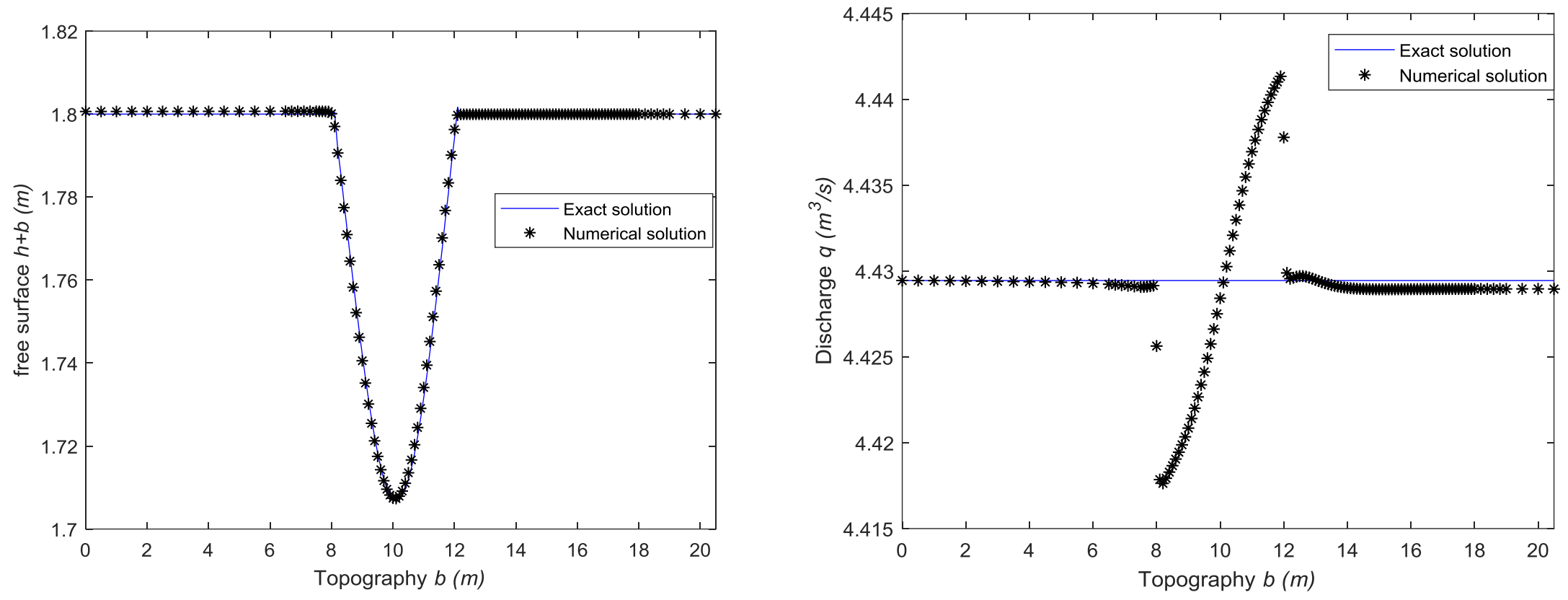

Fig 6 Fluvial flow: Comparison of free surfaces *h+b* (left) and discharge *q* (right) resulting from the new scheme represented by asterisk and the exact solution represented by line for a bottom topography *b*.

### 6.4 Transcritcal flow from fluvial regime to torrential regime

The free surface is initially constant with a height 0.13 *m* which submerges all the topography. Only a discharge of 0.6 $m^3/s$ is imposed at the left side. The time step is 0.01 *s*, and the total number of time steps 10000 which is long enough to ensure the steady state. The simulation result is depicted in Fig 7.

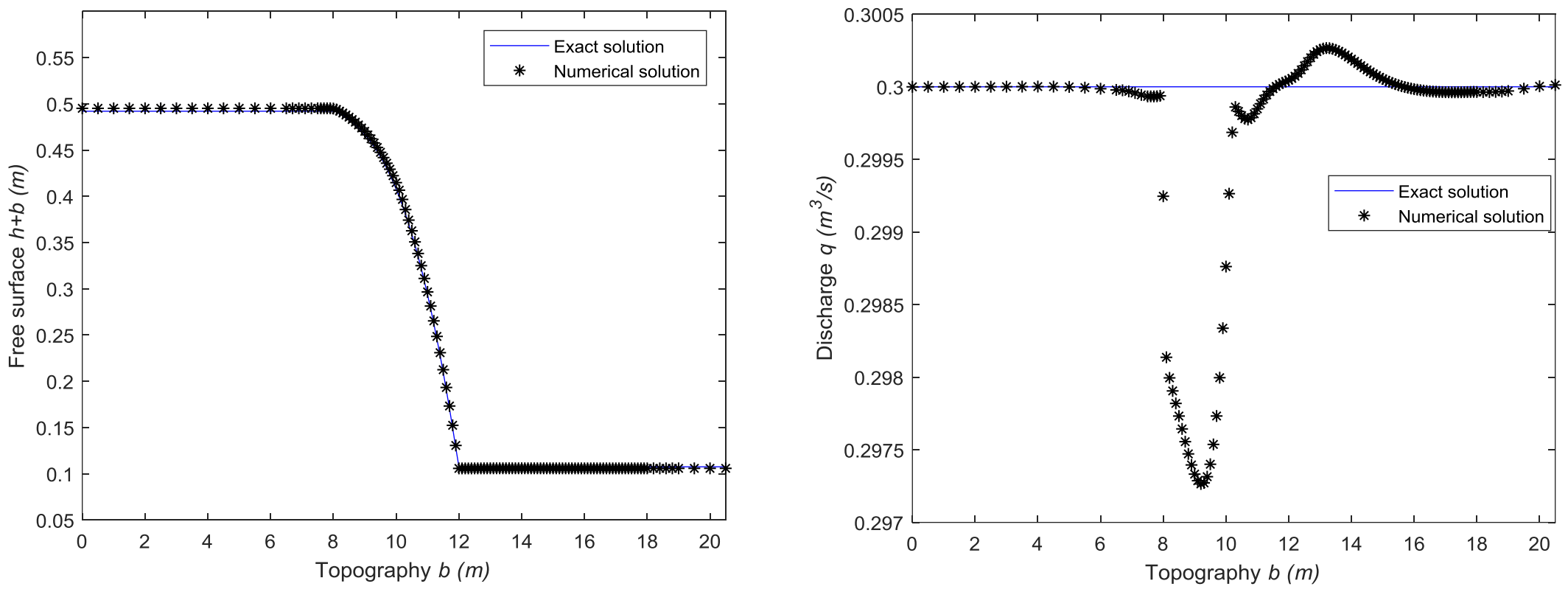

Fig 7 Transcritical flow: Comparison of free surfaces *h+b* (left) and discharge *q* (right) resulting from the new scheme represented by asterisk and the exact solution represented by line for a bottom topography *b*.

### 6.5 Transcritical flow with a hydraulic jump

The initial condition is the same as that in 6.3. A discharge of 0.18 $m^3/s$ is imposed at the left side and a water depth with the height 0.13 *m* imposed at the right side. The time step is 0.01 *s*, and the total number of time steps 30000 which is long enough to ensure the steady state. The simulation result is depicted in Fig 8.

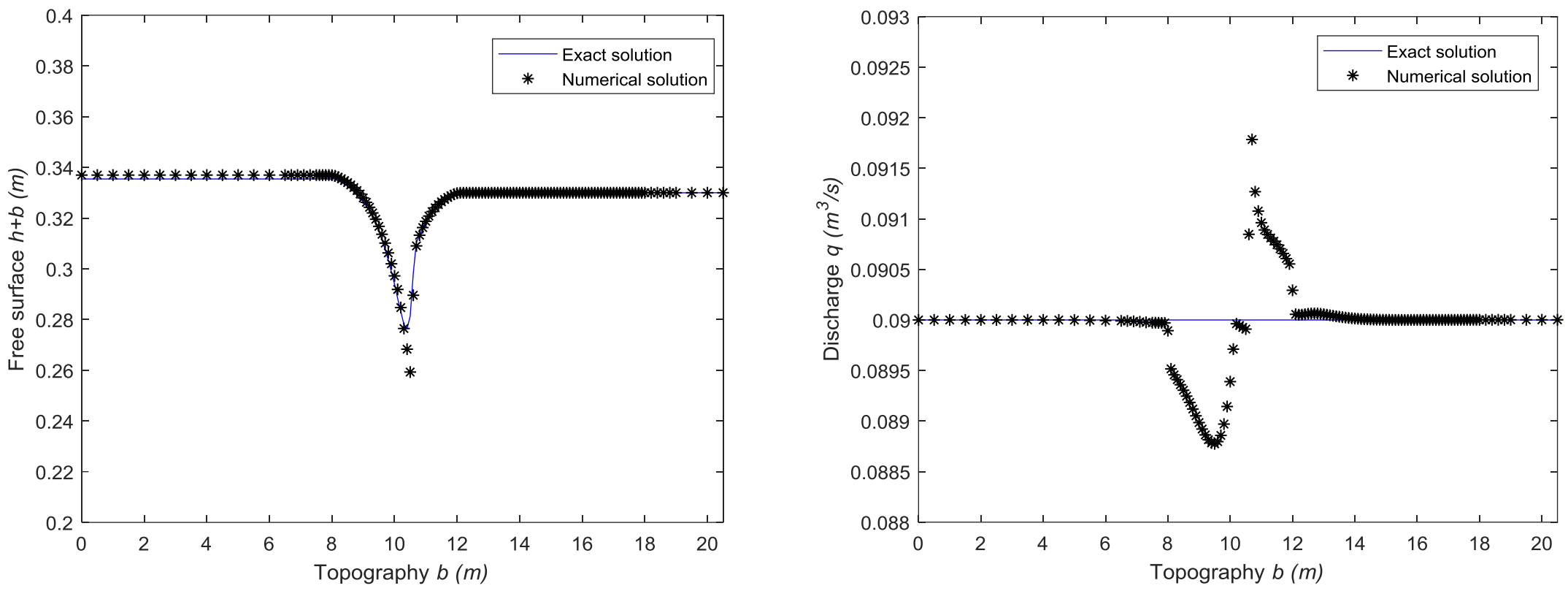

Fig 8 Flow with hydraulic jump: Comparison of free surfaces $h+b$ (left) and discharge $q$ (right) resulting from the new scheme represented by asterisk and the exact solution represented by line for a bottom topography $b$.

From Fig 6-8, the numerical result of our new scheme and analytic solution are in very good agreement. By looking carefully at the differences in free surface, it appears that the difference is extremely low. While for discharge, the maximum differences are 0.27%, 0.93% and 2.22%, respectively. The fluctuation in discharge from the new scheme is due to the numerical error which can be improved by densifying the computing mesh number. It is also clear that the fluctuation is tiny and the accuracy can be guaranteed.

## Conclusion

We have proposed a simple numerical scheme for the two dimensional Shallow-Water Equations (SWEs). This new scheme can ensure the well-balance condition and is proved to be accurate on several typical test cases: Fluvial regime, transcritical flow from fluvial regime to torrential regime and transcritical flow with a hydraulic jump.

## Acknowledgements

This study is one part of my final year's internship financially supported by *LNHE R&D, EDF*. Thanks is given to Dr Ung for his attentive explanation of his new numerical approach solving SWEs and my advisor Dr Goutal for her kindly guidance. Thanks is also given to Prof Delestre for his sincere suggestions.